\documentclass[journal]{IEEEtran}
\usepackage[numbers,sort&compress]{natbib}
\usepackage{amsmath,graphicx}
\usepackage{multirow}
\usepackage{booktabs}

\ifCLASSINFOpdf
\else
\fi
\begin{document}
%
\title{Granger Causality Analysis Based on Quantized Minimum Error Entropy Criterion}
%
%
%

\author{Badong~Chen,~\IEEEmembership{Senior Member,~IEEE,}
        Rongjin~Ma,
        Siyu~Yu,
        Shaoyi~Du,
        and~Jing~Qin,~\IEEEmembership{Member,~IEEE}
\thanks{This work was supported by 973 Program (No. 2015CB351703) and National NSF of China (No. 91648208, No. U1613219).}
\thanks{Badong~Chen (chenbd@mail.xjtu.edu.cn), Rongjin~Ma, Siyu~Yu and Shaoyi~Du are with the School of Electronic and Information Engineering,
Xi'an Jiaotong University, Xi'an 710049, Shaanxi, China.}
\thanks{Jing~Qin is with the Center of Smart Health, School of Nursing, Hong Kong Polytechnic University, Hongkong, China.}

}
%
%
\markboth{Journal of \LaTeX\ Class Files,~Vol.~14, No.~8, August~2015}%
{Shell \MakeLowercase{\textit{et al.}}: Bare Demo of IEEEtran.cls for IEEE Journals}
%
\maketitle
\begin{abstract}

Linear regression model (LRM) based on mean square error (MSE) criterion is widely used in Granger causality analysis (GCA), which is the most commonly used method to detect the causality between a pair of time series. However, when signals are seriously contaminated by non-Gaussian noises, the LRM coefficients will be inaccurately identified. This may cause the GCA to detect a wrong causal relationship. Minimum error entropy (MEE) criterion can be used to replace the MSE criterion to deal with the non-Gaussian noises. But its calculation requires a double summation operation, which brings computational bottlenecks to GCA especially when sizes of the signals are large. To address the aforementioned problems, in this study we propose a new method called GCA based on the quantized MEE (QMEE) criterion (GCA-QMEE), in which the QMEE criterion is applied to identify the LRM coefficients and the quantized error entropy is used to calculate the causality indexes. Compared with the traditional GCA, the proposed GCA-QMEE not only makes the results more discriminative, but also more robust. Its computational complexity is also not high because of the quantization operation. Illustrative examples on synthetic and EEG datasets are provided to verify the desirable performance and the availability of the GCA-QMEE.

\end{abstract}


%
\IEEEpeerreviewmaketitle

\section{Introduction}
%
%
%
%
\IEEEPARstart{G}{ranger} causality analysis (GCA) is one of the most commonly used methods to detect the causality between a pair of time series, which finds successful applications in various areas, such as economics \cite{paul2004causality, kar2011financial}, climate studies \cite{mosedale2006granger, pao2011multivariate}, genetics \cite{zhu2010characterizing}, and neuroscience \cite{roebroeck2005mapping, goebel2003investigating}. GCA is based on the linear regression model (LRM), so it is easy to understand and utilize \cite{kaminski2001evaluating}. However, the mean square error (MSE) criterion used in LRM takes only the second order moments of the errors into account. When signals are seriously contaminated by non-Gaussian noises, it becomes hard to accurately identify the LRM coefficients \cite{chen2016insights} and this may cause the traditional GCA to wrongly detect the causal relationships. The minimum error entropy (MEE) criterion in information theoretic learning (ITL) \cite{chen2016insights, erdogmus2006linear, Erdogmus2000Comparison, principe2010information, chen2013system, Erdogmus2003Convergence}, which takes higher order statistics of the errors into account, is generally superior to the MSE criterion in LRM identification especially when signals are contaminated by non-Gaussian noises, such as mixture Gaussian or Levy alpha-stable noises \cite{fofack1999tail}. But the calculation of the MEE based objective functions requires a double summation operation, which brings computational bottlenecks to GCA especially when the sizes of the signals are large. To address this issue, the quantized MEE (QMEE) criterion, a simplified version of the MEE criterion, was recently proposed in \cite{chen2017quantized}.

In this paper, we propose a new causality analysis method called GCA based on QMEE criterion (GCA-QMEE). The new method applies the QMEE to identify the LRM coefficients and uses the quantized error entropy to calculate the causality indexes. Compared with the traditional GCA, the proposed GCA-QMEE not only makes the results more discriminative, but also more robust. Because of the quantization operation, the computational complexity of GCA-QMEE is also not high.

The rest of the paper is organized as follows. Section 2 describes the original GCA method. Section 3 first introduces the QMEE criterion and its application to linear regression, then develops the GCA-QMEE method. In section 4, experiments are conducted to demonstrate the desirable performance and the availability of the proposed method. Finally, conclusion is given in section 5.

\section{Granger Causality Analysis}

In causality analysis, two vectors \emph{X} and \emph{Y} are considered as a pair of time series, which can be written as
\begin{equation}
\label{eq1}
\begin{aligned}
\textbf{X} &= \left[ {{x_1},{x_2}, \cdots, x{}_N} \right]^{T}\\
\textbf{Y} &= \left[ {{y_1},{y_2}, \cdots, y{}_N} \right]^{T}
\end{aligned}
\end{equation}
where \emph{N} is the size of the time series. Autoregressive (AR) models are used to fit the time series in Granger causality analysis (GCA) which are \cite{hesse2003use}
\begin{equation}
\label{eq2}
\begin{aligned}
{x_t} &= \sum\limits_{i = 1}^{{p_1}} {{w_{11,i}}{x_{t - i}}}  + {e_{11,t}}\\
{y_t} &= \sum\limits_{i = 1}^{{p_2}} {{w_{21,i}}{y_{t - i}}}  + {e_{21,t}}
\end{aligned}
\end{equation}
where ${p_1}$ and ${p_2}$ are the orders of the AR models, $\emph{e}_{11}$ and ${\emph{e}_{21}}$ are the errors, and ${\emph{w}_{11}}$ and ${\emph{w}_{21}}$ are the coefficients which can easily be solved by the least squares method based on the MSE criterion in linear regression. Under the MSE criterion, the cost function for estimating the coefficients is
\begin{equation}
\label{eq3}
J = \frac{1}{N}\sum\limits_{i = 1}^N {{e_i}}
\end{equation}
where $ \textbf{e} =[e_{1},e_{2},\cdots,e_{N}]^{T}$ are the error samples.

Vector autoregressive (VAR) models are also used to fit the time series in GCA, given by \cite{hesse2003use}
\begin{equation}
\label{eq4}
\begin{aligned}
{x_t} &= \sum\limits_{i = 1}^{{p_3}} {{w_{12,i}}{x_{t - i}}}  + \sum\limits_{j = 1 + {p_3}}^{2{p_3}} {{w_{12,j}}{y_{t - j + {p_3}}}}  + {e_{12,t}}\\
{y_t} &= \sum\limits_{i = 1}^{{p_4}} {{w_{22,i}}{y_{t - i}}}  + \sum\limits_{j = 1 + {p_4}}^{2{p_4}} {{w_{22,j}}{x_{t - j + {p_4}}}}  + {e_{22,t}}
\end{aligned}
\end{equation}
where ${p_3}$ and ${p_4}$ are the orders of the VAR models, ${\emph{w}_{12}}$ and ${\emph{w}_{22}}$ are the coefficients, and ${\emph{e}_{12}}$ and ${\emph{e}_{22}}$ are the error samples. If \emph{X} and \emph{Y} are independent, then ${\mathop{\rm Var}\nolimits} ({\emph{e}_{11}}) = {\mathop{\rm Var}\nolimits} ({\emph{e}_{12}})$ and ${\mathop{\rm Var}\nolimits} ({\emph{e}_{21}}) = {\mathop{\rm Var}\nolimits} ({\emph{e}_{22}})$, where ${\mathop{\rm Var}\nolimits} (\emph{e})$ denotes the variance of the error \emph{e}. Otherwise, the two equations don't hold. For example, if \emph{X} is the cause of \emph{Y}, then ${\mathop{\rm Var}\nolimits} ({\emph{e}_{21}}) > {\mathop{\rm Var}\nolimits} ({\emph{e}_{22}})$. Two Granger causality indexes can be computed by \cite{bressler2011wiener}
\begin{equation}
\label{eq5}
\begin{aligned}
{{\mathop{\rm F}\nolimits} _{X \to Y}} &= \log \frac{{{\mathop{\rm Var}\nolimits} ({\emph{e}_{21}})}}{{{\mathop{\rm Var}\nolimits} ({\emph{e}_{22}})}}\\
{{\mathop{\rm F}\nolimits} _{Y \to X}} &= \log \frac{{{\mathop{\rm Var}\nolimits} ({\emph{e}_{11}})}}{{{\mathop{\rm Var}\nolimits} ({\emph{e}_{12}})}}
\end{aligned}
\end{equation}

Clearly, we have ${{\mathop{\rm F}\nolimits} _{X \to Y}} \ge 0$ and ${{\mathop{\rm F}\nolimits} _{Y \to X}} \ge 0$. With the above indexes, the causality can be analyzed. Specifically, if ${{\mathop{\rm F}\nolimits} _{X \to Y}} > {{\mathop{\rm F}\nolimits} _{Y \to X}}$, then \emph{X} is the cause of \emph{Y}, or the information flowing from \emph{X} to \emph{Y} is more than that from \emph{Y} to \emph{X}; if ${{\mathop{\rm F}\nolimits} _{X \to Y}} < {{\mathop{\rm F}\nolimits} _{Y \to X}}$, then \emph{Y} is the cause of \emph{X} \cite{bressler2011wiener}.

\section{QMEE Based Granger Causality Analysis}

In the traditional GCA, the LRM coefficients may be inaccurately identified by the least squares method especially when signals are contaminated by non-Gaussian noises. To solve this problem, we use the QMEE criterion instead of the MSE criterion to estimate the LRM coefficients and propose a new causality analysis method called GCA-QMEE.

\subsection{LRM Identification Based on QMEE}

In ITL, Renyi's entropy of order $\alpha$ ($\alpha > 0, \alpha\neq1$) is widely used as a cost function, which is defined by \cite{Renyi1961On}
\begin{equation}
\label{eq6}
{H_ \alpha}(e) = \frac{1}{1-\alpha} \log \bigg [\int_e {{p^ \alpha}(e)de} \bigg ]
\end{equation}
where $p(e)$ denotes the probability density function (PDF) of the error variable \emph{e}, which is often estimated by the Parzen window approach \cite{silverman2018density}:
\begin{equation}
\label{eq7}
p(e) \approx \frac{1}{N}\sum\limits_{i = 1}^N {{G_\sigma }(e,{e_i})}
\end{equation}
where ${G_\sigma }(e, e_{i} )$ is the Gaussian kernel function with bandwidth $\sigma$:
\begin{equation}
\label{eq8}
{G_\sigma }(e,{e_i}) = \frac{1}{{\sqrt {2\pi } \sigma }}{\exp \bigg [ - \frac{{{{(e - {e_i})}^2}}}{{2{\sigma ^2}}} \bigg ]}
\end{equation}

In this work, without explicit mention we set $\alpha  = 2$. In this case, we have \cite{principe2010information}
\begin{equation}
\label{eq9}
{H_2}\left( e \right) \approx  - \log \left[ {\frac{1}{{{N^2}}}\sum\limits_{i = 1}^N {\sum\limits_{j = 1}^N {{G_{\sqrt 2 \sigma }}\left( {{e_i} , {e_j}} \right)} } } \right]
\end{equation}

There is a double summation operator in (9). Thus the computational cost for evaluating the error entropy is $O(N^2)$, which is expensive especially for large scale datasets. To reduce the computational complexity of the MEE criterion, we proposed an efficient quantization method in in a recent paper \cite{chen2017quantized}. Given \emph{N} error samples $\textbf{e}=[e_{1},e_{2},\cdots,e_{N}]$ and a quantization threshold $\epsilon$, the outputs of this quantization method are $Q(\textbf{e})=[Q(e_{1}),Q(e_{2}),\cdots,Q(e_{N})]$ and a codebook \textbf{C} containing \emph{M} real valued code words ($\emph{M}\ll\emph{N}$), where $Q(\cdot)$ denotes a quantization operator (See \cite{chen2017quantized, chen2012quantized, chen2013quantized} for the details of the quantization operator $Q(\cdot)$). If $\textbf{C}=[c_{1},c_{2},\cdots,c_{M}]$, then the quantized error entropy is
\begin{equation}
\label{eq10}
\begin{aligned}
H_2^Q(e)&=  - \log \left\{ {\frac{1}{{{N^2}}}\sum\limits_{i = 1}^N {\sum\limits_{j = 1}^N {{G_{\sqrt 2 \sigma }}\left[ {{e_i} , Q\left( {{e_j}} \right)} \right]} } } \right\}\\ &=  - \log \left[ {\frac{1}{{{N^2}}}\sum\limits_{i = 1}^N {\sum\limits_{m = 1}^M {{A_m}{G_{\sqrt 2 \sigma }}\left( {{e_i} , {c_m}} \right)} } } \right]\\
&=  - \log \left[ I_2^Q(e) \right]
\end{aligned}
\end{equation}
where ${{A_m}}$ is the number of the error samples that are quantized to the code word ${{c_m}}$, and $I_2^Q(e)$ is called the quantized information potential \cite{chen2017quantized}. Under the QMEE criterion, the optimal hypothesis can thus be solved by minimizing the quantized error entropy $H_2^Q(e)$. Clearly, minimizing $H_2^Q(e)$ is equivalent to maximizing $I_2^Q(e)$.

Consider the LRM in which $e_{i}=y_{i}-\textbf{w}^{T}\textbf{x}_{i}$ with $\textbf{w}$ being the coefficient vector to be estimated.
Taking the gradient of $I_2^Q(e)$ with respect to $\textbf{w}$, we have
\begin{equation}
\label{eq11}
\begin{array}{l}
\frac{{\partial I_2^Q(e)}}{{\partial {\textbf{w}}}} = {\frac{1}{{{N^2}}}\sum\limits_{i = 1}^N {\sum\limits_{m = 1}^M {{A_m}\frac{{\partial {G_{\sqrt 2 \sigma }}\left( {{e_i} , {c_m}} \right)}}{{\partial {\textbf{w}}}}} } } \\
 =  {\tau \sum\limits_{i = 1}^N {\sum\limits_{m = 1}^M {{A_m}{G_{\sqrt 2 \sigma }}\left( {{e_i} , {c_m}} \right)\left( {{y_i} - {\textbf{w}^T}{\textbf{x}_i} - {c_m}} \right){\textbf{x}_{i}}} } }\\
 =  {\tau \sum\limits_{i = 1}^N {\sum\limits_{m = 1}^M {{A_m}{G_{\sqrt 2 \sigma }}\left( {{e_i} , {c_m}} \right)\left( {{y_i} - {c_m}} \right){\textbf{x}_i}} } }\\
 + { \tau\left[ {\sum\limits_{i = 1}^N {\sum\limits_{m = 1}^M {{A_m}{G_{\sqrt 2 \sigma }}\left( {{e_i} , {c_m}} \right){\textbf{x}_{i}}{\textbf{x}_i}^T} } } \right]\textbf{w}}
\end{array}
\end{equation}
where $\tau = \frac{1}{{{N^2}{{(\sqrt 2 \sigma )}^2}}}$. Setting $\frac{{\partial I_2^Q(e)}}{{\partial {\textbf{w}}}} = 0$, a fixed point equation of the LRM coefficients can be obtained as
\begin{equation}
\label{eq12}
\textbf{w} = {\textbf{V}^{ - 1}}\textbf{U}
\end{equation}
where $\textbf{U} = \sum\limits_{i = 1}^N {\sum\limits_{m = 1}^M {{A_m}{G_{\sqrt 2 \sigma }}\left( {{e_i} , {c_m}} \right)\left( {{y_i} - {c_m}} \right){\textbf{x}_i}} } $ and $\textbf{V} = {\sum\limits_{i = 1}^N {\sum\limits_{m = 1}^M {{A_m}{G_{\sqrt 2 \sigma }}\left( {{e_i} , {c_m}} \right){\textbf{x}_{i}}{\textbf{x}_i}^T} } } $. After reaching the steady state by multiple fixed-point iterations ($\textbf{w}_{k}=(\textbf{V}^{-1}\textbf{U})|_{\textbf{w}_{k-1}}, k=1,2,\cdots,K$), the fixed-point solution of the coefficients can be obtained.

Here we present an illustrative example to compare the performance of the MSE, MEE and QMEE criterions. Consider the following linear system:
\begin{equation}
\label{eq13}
{y_i} = {\textbf{w}^{{*^T}}}{\textbf{x}_i} + {\varphi _i}
\end{equation}
where ${\textbf{w}^*}=[2,1]^{T}$, $\textbf{x}_{i}$ is assumed to be uniformly distributed over $[-2,2]\times[-2,2]$, and  $\varphi_{i}$ is a non-Gaussian noise drawn from:

Case 1) symmetric Gaussian mixture density: $f(x) = 0.5 \times \frac{1}{{\sqrt {2\pi }  \times 1}}{\exp[ - \frac{{{{\left( {x - 4} \right)}^2}}}{{2 \times {1^2}}}]} + 0.5 \times \frac{1}{{\sqrt {2\pi }  \times 1}}{\exp[ - \frac{{{{\left( {x + 4} \right)}^2}}}{{2 \times {1^2}}}]}$;

Case 2) asymmetric Gaussian mixture density: $f(x) = 0.6 \times \frac{1}{{\sqrt {2\pi }  \times 1}}{\exp[ - \frac{{{{\left( {x - 3} \right)}^2}}}{{2 \times {1^2}}}]} + 0.4 \times \frac{1}{{\sqrt {2\pi }  \times 1}}{\exp[ - \frac{{{{\left( {x + 5} \right)}^2}}}{{2 \times {1^2}}}]}$;

Case 3) Levy alpha-stable distribution with characteristic exponent ($0<\alpha\leq2$), skewness ($-1\leq\beta\leq1$), scale parameter ($0<\gamma<\infty$) and location parameter ($-\infty<\delta<\infty$) \cite{fofack1999tail}, where $[\alpha,\beta,\gamma,\delta]=[1.3,0,0.4,0]$.

The root mean squared error (RMSE) is employed to measure the performance, computed by
\begin{equation}
\label{eq14}
RMSE = \sqrt {\frac{1}{2}{{\left\| {{\textbf{w}^*} - \widehat \textbf{w}} \right\|}^2}}
\end{equation}
where $\textbf{w}^{*}$ and $\widehat \textbf{w}$ denote the target and the estimated weight vectors respectively.

For the MSE criterion, there is a closed-form solution, so no iteration is needed. For other two criterions, the fixed-point iteration is used to solve the model. In the simulations, the size of samples is $N = 500$, the iteration number is $K = 100$, the Gaussian kernel bandwidth is $\sigma = 0.5$ and the quantization threshold is set at $\epsilon = 0.4$. The mean $\pm$ deviation results of the RMSE over 100 Monte Carlo runs are presented in Table I. From Table I, we observe i) the MEE and QMEE criterions can significantly outperform the traditional MSE criterion; ii) the QMEE criterion can achieve almost the same (or even better) performance as the original MEE criterion.
\begin{table}[!htbp]
\centering
\caption{Mean $\pm$ deviation results of RMSE over 100 Monte Carlo runs}\label{RMSE}
\begin{tabular}{cccc}
  \toprule
  & Case 1) & Case 2) & Case 3) \\
  \hline
  MSE & 0.1437 $\pm$ 0.0755 & 0.1454 $\pm$ 0.0733 & 0.3297 $\pm$ 1.6278 \\
  MEE & 0.0414 $\pm$ 0.0232 & 0.0413 $\pm$ 0.0224 & 0.0216 $\pm$ 0.0107 \\
  QMEE& 0.0436 $\pm$ 0.0232 & 0.0428 $\pm$ 0.0237 & 0.0215 $\pm$ 0.0106 \\
  \toprule
\end{tabular}
\end{table}

Fig.1 shows the training time of the QMEE and MEE with increasing sample size. Obviously, compared with the MEE criterion, the computational complexity of the QMEE criterion is very low.
\begin{figure}[htb]
  \includegraphics[width=7cm]{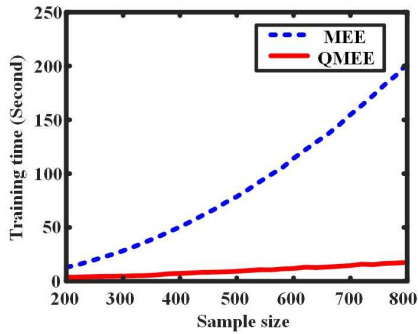}
  \caption{Training time with different sizes of samples.}\label{fig1}
\end{figure}

\subsection{GCA Based on QMEE}

When signals are seriously contaminated by non-Gaussian noises, the traditional GCA may not work well. We propose a new causality analysis method called GCA-QMEE to solve this problem, in which the QMEE criterion is applied to identify (by fixed-point iterations) the LRM coefficients, namely, the coefficients of the AR and VAR models in (2) and (4). The quantized error entropy is used to calculate the causality indexes:
\begin{equation}
\label{eq15}
\begin{aligned}
{{\mathop{\rm F}\nolimits} _{X \to Y}} &= {{H_2^Q({e_{21}})}} - {{H_2^Q({e_{22}})}} = \log \left[ {\frac{{I_2^Q({e_{22}})}}{{I_2^Q({e_{21}})}} } \right]\\
{{\mathop{\rm F}\nolimits} _{Y \to X}} &= {{H_2^Q({e_{11}})}} - {{H_2^Q({e_{12}})}} = \log \left[ {\frac{{I_2^Q({e_{12}})}}{{I_2^Q({e_{11}})}} } \right]
\end{aligned}
\end{equation}
where $H_2^Q({e_{11}})$, $H_2^Q({e_{21}})$, $H_2^Q({e_{12}})$ and $H_2^Q({e_{22}})$ are the trained quantized error entropies of the LRMs, and $I_2^Q({e_{11}})$, $I_2^Q({e_{21}})$, $I_2^Q({e_{12}})$ and $I_2^Q({e_{22}})$ are the trained quantized information potentials after convergence. The orders (or the embedding dimensions) of the LRMs in our approach can be determined by the following Bayesian Information Criterion (BIC) \cite{kass1995reference}:
\begin{equation}
\label{eq16}
p= \mathop {\min \arg }\limits_{1 \le i \le {p_{\max }}} \left[ {N \log \left( H_2^Q({e}) \right) + i \log \left( N \right)} \right]
\end{equation}
where $p_{\max }$ is the largest embedding dimension we set.

\section{Experimental Results}

In this section, we test the proposed GCA-QMEE on two datasets. One is a synthetic dataset, and the other is an EEG dataset.

\subsection{Synthetic Dataset}

Suppose the time series \emph{X} and \emph{Y} are generated by
\begin{equation}
\label{eq17}
\begin{aligned}
{x_t} &= {\varphi _t}\\
{y_t} &= {x_{t - 1}} + {\psi _t}
\end{aligned}
\end{equation}
where $\varphi_{t}$ denotes a noise following the uniform distribution over $[-2,2]$, and $\psi_{t}$ is a non-Gaussian noise drawn from the three PDFs in the illustrative example of the previous section. In this example, obviously, \emph{X} is the cause of \emph{Y} but not vice versa.

We compare the performance of several GCA methods: GCA-MSE, GCA-MEE and GCA-QMEE, where GCA-MSE represents the traditional GCA, and GCA-MEE corresponds to the GCA-QMEE with quantization threshold $\epsilon=0$. Since \emph{X} is the cause of \emph{Y}, we define the following discrimination index to measure the performance of causality detection:
\begin{equation}
\label{eq18}
\rho = \frac{{{\mathop{\rm F}\nolimits} _{X \to Y}}-{{\mathop{\rm F}\nolimits} _{Y \to X}}}{{{\mathop{\rm F}\nolimits} _{X \to Y}}}
\end{equation}
Clearly, if $\rho>0$, then ${{\mathop{\rm F}\nolimits} _{X \to Y}}>{{\mathop{\rm F}\nolimits} _{Y \to X}}$, and thus the causal relationship is correctly detected; if $\rho<0$, then the causal relationship is wrongly detectly. The bigger the $\rho$ is, the more discriminative the causality analysis result is (i.e. the difference between ${{\mathop{\rm F}\nolimits} _{X \to Y}}$ and ${{\mathop{\rm F}\nolimits} _{Y \to X}}$ is more significant). The parameter settings are $N = 500, K = 100, \sigma = 0.5, \epsilon = 0.4$ and $p_{\max }=10$. The causality indexes over 100 Monte Carlo experiments are presented in TABLE II. From TABLE II, we observe: i) all methods can correctly detect that \emph{X} is the cause of \emph{Y}; ii) the GCA-QMEE and the GCA-MEE can significantly outperform the GCA-MSE in terms of the discrimination index.
\begin{table}[!htbp]
\centering
\caption{Causality indexes of different methods}\label{Synthetic_Dataset}
\begin{tabular}{p{0.5cm}p{0.5cm}p{2cm}p{2cm}p{2cm}}
  \toprule
   & & GCA-MSE & GCA-MEE & GCA-QMEE \\
  \hline
   \multirow{3}*{Case 1)}& $\texttt{F}_{X\rightarrow Y}$ & 0.0807 $\pm$ 0.0243 & 0.3842 $\pm$ 0.0299 & 0.3819 $\pm$ 0.0304 \\
   & $\texttt{F}_{Y\rightarrow X}$ & 0.0032 $\pm$ 0.0055 & 0.0009 $\pm$ 0.0011  & 0.0011 $\pm$ 0.0045 \\
   & $\rho$ & 0.9588 $\pm$ 0.0783 & 0.9977 $\pm$ 0.0030 & 0.9973 $\pm$ 0.0118 \\
   \hline
   \multirow{3}*{Case 2)}& $\texttt{F}_{X\rightarrow Y}$ & 0.0782 $\pm$ 0.0234 & 0.3787 $\pm$ 0.0303 & 0.3755 $\pm$ 0.0295 \\
   & $\texttt{F}_{Y\rightarrow X}$ & 0.0035 $\pm$ 0.0071 & 0.0011 $\pm$ 0.0016 & 0.0004 $\pm$ 0.0047 \\
   & $\rho$ & 0.9512 $\pm$ 0.0925 & 0.9970 $\pm$ 0.0043 & 0.9990 $\pm$ 0.0129 \\
    \hline
   \multirow{3}*{Case 3)}& $\texttt{F}_{X\rightarrow Y}$ & 0.1993 $\pm$ 0.1575 & 0.5828 $\pm$ 0.0298 & 0.5773 $\pm$ 0.0311 \\
   & $\texttt{F}_{Y\rightarrow X}$ & 0.0027 $\pm$ 0.0040 & 0.0008 $\pm$ 0.0010 & 0.0004 $\pm$ 0.0038 \\
   & $\rho$ & 0.9284 $\pm$ 0.3865 & 0.9986 $\pm$ 0.0017 & 0.9993 $\pm$ 0.0064 \\
   \toprule
\end{tabular}
\end{table}

To further show the robustness of our proposed method, we evaluate the relative variation ratio (RVR) of the causality index ${{\mathop{\rm F}\nolimits} _{X \to Y}}$ computed by GCA-QMEE, GCA-MEE and GCA-MSE in Levy alpha-stable noise ($\psi_{t}$). Here the RVR is defined by
\begin{equation}
\label{eq19}
\xi(\alpha)  = \left| {\frac{{{{\mathop{\rm F}\nolimits}_{X \to Y}(\alpha)} - {{\mathop{\rm F}\nolimits}_{X \to Y}(2)}}}{{{{\mathop{\rm F}\nolimits}_{X \to Y}(2)}}}} \right|, 0<\alpha\leq2
\end{equation}
where ${{\mathop{\rm F}\nolimits}_{X \to Y}(\alpha)}$ denotes the causality index ${{\mathop{\rm F}\nolimits} _{X \to Y}}$ obtained when signals are contaminated by Levy alpha-stable distribution with characteristic exponent $\alpha$ $(0<\alpha\leq2)$, and ${{\mathop{\rm F}\nolimits}_{X \to Y}(2)}$ corresponds to the causality index obtained in Gaussian noises. Clearly, $\xi(\alpha)$ measures the change when noise $\psi_{t}$ changes from the Gaussian distribution ($\alpha=2$) to non-Gaussian distribution ($\alpha<2$). Obviously, the smaller the $\xi(\alpha)$ is, the more robust the causality detection result is (that is, the change of the causality index is small when noise distribution is changing). The RVRs averaged over 50 Monte Carlo runs are shown in Fig. 2, where the parameters of the Levy alpha-state distribution are $[\alpha,0,0.4,0]$, and $\alpha$ varies from 2.0 to 0.5 with step 0.05. It is worth noting that when $\alpha$ becomes smaller, the noises will be more impulsive. From Fig. 2, one can see that the RVR of the traditional GCA method changes a lot when $\alpha$ changes from 2.0 to 0.5, while the RVRs of the GCA-QMEE and GCA-MEE change very little. This confirms that the GCA-QMEE and GCA-MEE are more robust than the traditional GCA.
\begin{figure}[htb]
  \includegraphics[width=7cm]{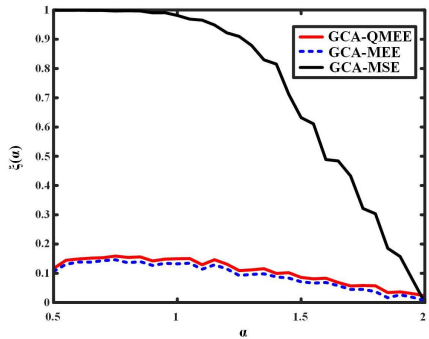}
  \caption{Relative variation ratio of the causality from \emph{X} to \emph{Y} versus $\alpha$.}\label{fig2}
\end{figure}

\subsection{EEG Dataset}

Now we apply the GCA-QMEE to analyze the dataset IIb of BCI competition IV \cite{leeb2008bci}. The dataset are recorded from nine subjects with three electrodes (C3, Cz and C4). The subjects are right handed, and have normal or corrected-to-normal vision. There are two different motor imagery (MI) tasks, namely, left-hand MI and right-hand MI. The sample frequency is 250Hz. The EEG signals are already band-pass filtered between 0.5Hz and 100Hz with a 50Hz notch filter enabled. For each subject, we only use the first three sessions. Hence, there are 27 sessions for our analysis. Each session contains some trails. For each session, we extract all the valid trail data, then average the data as a set of EEG time series. We analyze the causalities among C3, Cz and C4 by the GCA-QMEE. The parameter settings are $K = 100, \sigma = 0.5, \epsilon = 0.05$ and $p_{\max }=20$. After calculation, we get 27 causality indexes among C3, Cz and C4. The results are averaged in TABLE III.
\begin{table}[!htbp]
\centering
\caption{Causality indexes among C3, Cz and C4 for left-hand MI and right-hand MI}\label{EEG Dataset}
\begin{tabular}{ccc}
  \toprule
  & Left-hand MI & Right-hand MI \\
  \hline
  $\texttt{F}_{C3\rightarrow Cz}$& 0.0006 & 0.0015 \\
  $\texttt{F}_{Cz\rightarrow C3}$& 0.0038 & 0.0037 \\
  $\texttt{F}_{C3\rightarrow C4}$& 0.0014 & 0.0026 \\
  $\texttt{F}_{C4\rightarrow C3}$& 0.0024 & 0.0018 \\
  $\texttt{F}_{C4\rightarrow Cz}$& 0.0009 & 0.0015 \\
  $\texttt{F}_{Cz\rightarrow C4}$& 0.0039 & 0.0047 \\
  \toprule
\end{tabular}
\end{table}

From Table III, we observe: a) all causality indexes are greater than 0, indicating that there are several bidirectional causalities among C3, Cz and C4 during MI; b) causality indexes from Cz to C3/C4 are greater than those from C3/C4 to Cz during MI; c) causality index from C4 to C3 is larger than that from C3 to C4 during left-hand MI, and causality index from C3 to C4 is larger than that from C4 to C3 during right-hand MI; d) the causality from Cz to C4 during right-hand MI is larger than that from Cz to C3 during left-hand MI, and the causality from C3 to C4 during right-hand MI is larger than that from C4 to C3 during left-hand MI, which demonstrate the influence of the brain asymmetry of right-handedness on effective connectivity networks. These results validate the previous findings \cite{Chen2009Evaluation, gao2011evaluation, hu2014causality, hu2016comparison}.

\section{Conclusion}

In this work, we proposed a new causality analysis method called Granger causality analysis (GCA) based on the quantized minimum error entropy (QMEE) criterion (GCA-QMEE), in which the QMEE criterion is applied to identify the LRM coefficients and the quantized error entropy is used to calculate the causality indexes. Compared with the traditional GCA, the proposed GCA-QMEE not only makes the results more discriminative, but also more robust. Its computational complexity is also not high because of the quantization operation. Experimental results with synthetic and EEG datasets have been provided to confirm the desirable performance of the GCA-QMEE.

\ifCLASSOPTIONcaptionsoff
  \newpage
\fi

\bibliographystyle{IEEEtran}
\bibliography{GCA_QMEE}

\end{document}